\documentstyle[12pt,epsf,epsfig]{article}  

\def\li2{{\rm Li}_2}

\def\roughly#1{\,\,\raise.3ex\hbox{$#1$\kern-.75em\lower1ex\hbox{$\sim$}}\,\,}

\def \lsim{\mathrel{\vcenter
     {\hbox{$<$}\nointerlineskip\hbox{$\sim$}}}}

\def\fo{\hbox{{1}\kern-.25em\hbox{l}}}

\def\bea{\begin{eqnarray}}
\def\eea{\end{eqnarray}}
\def\beq{\begin{equation}}
\def\eeq{\end{equation}}
\def\eq{\end{equation}}
\def\to{\rightarrow}

\def\bsg{\ifmmode B\to X_s\gamma\else $B\to X_s\gamma$\fi}
\def\bsll{\ifmmode B\to X_s\ell^+\ell^-\else $B\to X_s\ell^+\ell^-$\fi}
\def\bstt{\ifmmode B\to X_s\tau^+\tau^-\else $B\to X_s\tau^+\tau^-$\fi}
\def\shat{\ifmmode \hat{s}\else $\hat{s}$\fi}

\newcommand{\newc}{\newcommand}

\newc{\lcal}{\int {\cal L}dt}
 
\newc{\LSP}{{\chi^0_1}}
\newc{\stauR}{{\tilde \tau_R}}
\newc{\stau}{{\tilde \tau_1}}
\newc{\mstop}{m_{\tilde{t}}}
\newc{\mHpm}{m_{H^\pm}}
\newc{\ie}{{\it i.e.}}          
\newc{\etal}{{\it et al.}}
\newc{\eg}{{\it e.g.}}          
\newc{\kev}{\hbox{\rm\,keV}}            
\newc{\mev}{\hbox{\rm\,MeV}}            
\newc{\gev}{\hbox{\rm\,GeV}}            
\newc{\tev}{\hbox{\rm\,TeV}}
\newc{\xpb}{\hbox{\rm\, pb}}
\newc{\xfb}{\hbox{\rm\, fb}}

\newc{\mtop}{m_t}
\newc{\mbot}{m_b}
\newc{\mz}{m_Z}
\newc{\mw}{M_W}
\newc{\alphasmz}{\alpha_s(m_Z^2)}
\newc{\swsq}{\sin^2\theta_W}
\newc{\tw}{\tan\theta_W}
\newc{\cw}{\cos\theta_W}
\newc{\sw}{\sin\theta_W}
\newc{\BR}{\hbox{\rm BR}}
\newc{\zbb}{Z\to b\bar}
\newc{\Gb}{\Gamma (Z\to b\bar b)}
\newc{\Gh}{\Gamma (Z\to \hbox{\rm hadrons})}
\newc{\rbsm}{R_b^\hbox{\rm sm}}
\newc{\rbsusy}{R_b^\hbox{\rm susy}}
\newc{\drb}{\delta R_b}

\newc{\sgn}{\mbox{sgn}}
\newc{\tbeta}{\tan\beta}
\newc{\uL}{{\tilde u_L}}
\newc{\uR}{{\tilde u_R}}
\newc{\cL}{{\tilde c_L}}
\newc{\cR}{{\tilde c_R}}
\newc{\tL}{{\tilde t_L}}
\newc{\tR}{{\tilde t_R}}
\newc{\dL}{{\tilde d_L}}
\newc{\dR}{{\tilde d_R}}
\newc{\sL}{{\tilde s_L}}
\newc{\sR}{{\tilde s_R}}
\newc{\bL}{{\tilde b_L}}
\newc{\bR}{{\tilde b_R}}
\newc{\eL}{{\tilde e_L}}
\newc{\eR}{{\tilde e_R}}
\newc{\mhp}{m_{H^\pm}}
\newc{\mhalf}{m_{1/2}}
\newc{\emt}{{e/\mu /\tau}}

\newc{\bW}{{\bar W}}
\newc{\bB}{{\bar B}}
\newc{\eps}{{\epsilon}}

\def\lappeq{\mathrel{\rlap{\raise.5ex\hbox{$<$}}
{\lower.5ex\hbox{$\sim$}}}}
\def\gappeq{\mathrel{\rlap {\raise.5ex\hbox{$>$}}
{\lower.5ex\hbox{$\sim$}}}}

\begin{document}

\baselineskip=18pt



 


\vspace{20pt}
\font\cmss=cmss10 \font\cmsss=cmss10 at 7pt 

\begin{flushright}
CERN-TH/2003-259\\
UAB-FT-554
\end{flushright}


\hfill

\vspace{20pt}

\begin{center}
{\Large \textbf
{Weakly coupled Higgsless theories\\
 and\\
precision electroweak tests\\}}
\end{center}

\vspace{6pt}

\begin{center}
\textsl{Riccardo Barbieri$\, ^a$, Alex Pomarol$\, ^b$ 
and Riccardo Rattazzi$\, ^c$} \vspace{20pt}

$^{a}$\textit{Scuola Normale Superiore and INFN, Pisa, Italy}

$^{b}$\textit{IFAE, Universitat Aut{\`o}noma de Barcelona,
08193 Bellaterra (Barcelona), Spain }

$^{c}$\textit{Theory Division, CERN, CH-1211 Geneva 23, Switzerland}
\end{center}

\vspace{12pt}

\begin{center}
\textbf{Abstract }
\end{center}

\vspace{4pt} {\small \noindent 
In 5 dimensions the electroweak symmetry can be broken by boundary conditions, 
leading to a new type of Higgsless theories. 
These could in principle improve on the 4D case by extending the
perturbative domain to energies higher than $4 \pi v$ and 
by allowing a better fit to the electroweak precision tests. 
Nevertheless, it is unlikely that both these improvements can be achieved,
as we show by discussing these problems in an explicit model.}

\vfill\eject 
\noindent


\section{Introduction}
Higgsless theories of the electroweak interactions do not appear to allow an acceptable description 
of the ElectroWeak Precision Tests (EWPT) \cite{lep}. At least, since they become strongly interacting at an
 energy of about $4 \pi v$, where $v$ is the Higgs vacuum expectation value, the calculability of the 
precision observables is limited. Furthermore, when an estimate can be made under suitable assumptions \cite{pt1}, 
the ultraviolet (UV)
 contribution to the parameter $\eps_3$ \cite{alba} 
(or to the parameter $S$ \cite{pt2}) 
is positive, which, added to the infrared
 (IR) piece cut off at some high scale $\sim 4\pi v$, 
makes it essentially impossible to fit the EWPT, 
independently of the value of  $\eps_1$ (or  of the $T$ parameter).
 
Breaking the electroweak symmetry by boundary conditions on an extra dimension may however 
be a new twist of the problem (see for example refs.~\cite{cdh,cgmpt}). 
The specific reason for this statement is the following. It is well
known that a gauge group $G$ can be broken by boundary conditions at compactification down to a subgroup $H$.
In this situation the vectors in $H$ have a Kaluza-Klein (KK) tower starting with a massless 4D mode, while
the lightest state in $G/H$ has mass $1/R$, where $R$ is the radius of compactification. A feature
of gauge symmetry breaking by boundary conditions, 
for example on orbifolds, is that in general it does not 
introduce new physical scales associated to a 
strongly interacting regime in the theory. For instance, using naive dimensional analysis (NDA)\cite{nda,nda2},
a five dimensional gauge theory will become
strongly coupled at a scale $\Lambda\sim 24\pi^3/g_5^2$, whether it is broken by boundary conditions or not.
Now the potentially interesting fact is that, if we interpret the 
lightest KK modes in $G/H$ as the $W$ and $Z$,  the 5D cut-off $\Lambda$ is written in terms of 4D
quantities as $\Lambda\sim 12 \pi^2 m_W/g_4^2$ (we have used $1/g_4^2=2\pi R/g_5^2$ and $m_W\sim 1/R$).
Compared to  the cut-off $4\pi m_W/g_4\sim 1$ TeV 
of a 4D Higgsless theory, the cut-off of the 5D Higgsless theory is
a factor $3\pi/g_4$ bigger: $\Lambda\sim 10$ TeV. To summarize, what happens 
physically is the following: an appropriate tower of KK states may play the role of the normal 4D 
Higgs boson in preventing the relevant amplitudes from exceeding the unitarity bounds up to an energy
scale $\Lambda$ which is not arbitrarily large but can be well above the cut-off $4 \pi v$ of a 4D Higgsless 
theory. This is why we call them weakly interacting Higgsless theories.  
 
One may do the same power counting exercise for a D-dimensional gauge theory compactified
down to 4-dimensions. In that case the cut-off is $\Lambda^{D-4}\sim (4\pi)^{D/2} \Gamma(D/2)/g_D^2$,
while the 4D gauge coupling is $1/g_4^2=(2\pi R)^{D-4}/g_D^2$. If we identify $m_W\sim 1/R$ we can then write
$\Lambda=m_W[16\Gamma(D/2)\pi^{(4-D/2)}/g_4^2]^{1/(D-4)}$, from which we conclude that by going to
$D>5$ we do not actually increase the cut-off. For $D=6$ the cut-off is again roughly $4\pi v$,
while for $D\to \infty$ it becomes $\sim m_W$. This result is intuitively clear: with a large number of extra 
dimensions, in order to keep $g_4$ finite, 
the radius of compactification should be right at the cut-off. 
We therefore stick to $D=5$.

The next obvious problem is to see how these 5D Higgsless theories can do in describing the EWPT,
 which is the purpose of this paper
\footnote{For a related discussion, see ref.~\cite{nomura}. }. 
We do this by analysing a specific model designed to keep 
under control the effects of the breaking of custodial isospin, so that we can focus on the 
effects of $\eps_3$ only. We shall also comment on the likely general validity of our conclusions.

\section{The model}

Motivated by the simple argument given in the introduction we consider a 5D gauge theory
compactified on $S_1/Z_2$.  Since we want calculable (and small) custodial symmetry breaking effects,
we must separate in 5D the sectors that break the electroweak symmetry from  those that break the custodial symmetry.
A necessary requirement to achieve this property is the promotion of custodial symmetry to a 
gauge symmetry. Then the minimal model with a chance of giving a realistic phenomenology
has gauge group $G=SU(2)_L\times SU(2)_R\times U(1)_{B-L}$. Following refs.~\cite{adms,cgpt}
we break $G\to SU(2)\times U(1)_Y$ at one boundary, $y=0$, and $G\to SU(2)_{L+R}\times U(1)_{B-L}$
at the other boundary, $y=\pi R$. 
Then the only surviving gauge symmetry below the compactification scale
is $U(1)_Q$. We achieve the breaking by explicitly adding mass terms for the broken generator vector
at the boundary and by sending this mass to infinity
\footnote{Even if  the boundary mass is kept to its natural value, $M\sim\Lambda$,
this will only modify our calculations by $ O(M_L/(\Lambda^2 R))$ effects.}.
The most general lagrangian up to two derivatives
is then
\beq
{\cal L}= {\cal L}_5+ \delta(y){\cal L}_0+\delta(y-\pi R){\cal L}_\pi\, ,
\eeq
with
\bea
{\cal L}_5&=& -\frac{M_L}{4} (W_L^I)_{MN}(W_L^I)^{MN}- \frac{M_R}{4} (W_R^I)_{MN}(W_R^I)^{MN}-
\frac{M_B}{4} B_{MN}B^{MN}\, ,\ \ \ \ (I=1,2,3)\nonumber\\
{\cal L}_0&=&-\frac{1}{4g^2} 
(W_L^I)_{\mu\nu}(W_L^I)^{\mu\nu}-\frac{1}{4{g'}^2} B_{\mu\nu}B^{\mu\nu}-\frac{M^2}{2}
\left[(W^+_R)_\mu(W^-_R)^\mu+(W^3_{R\, \mu}-B_\mu)(W^{3\, \mu}_R-B^\mu)\right]\, ,\nonumber\\
{\cal L}_\pi&=&-\frac{Z_W}{4} (W_L^I)_{\mu\nu}(W_L^I)^{\mu\nu}-\frac{Z_B}{4} B_{\mu\nu}B^{\mu\nu}
-\frac{M^2}{2}(W^I_{R\, \mu}-W^I_{L\, \mu})(W^{I\, \mu}_R-W^{I\, \mu}_L)\, .
\eea
In the limit $M\to \infty$ at $y=0$ we have $W^+_R=0$ and $ W^3_R=B$, then the kinetic terms in ${\cal L}_0$ 
are truly the most general ones. A similar comment applies at $y=\pi R$.

Notice that, although we have added extra operators at the boundaries, the UV cut-off of this theory is
still determined by the bulk gauge couplings. It is intuitively clear why the large mass terms at the boundaries
do not lower the cut-off: they can be viewed as originating from spontaneous gauge symmetry breaking by 
a $\sigma$-model sector  with typical scale $\sim M$. Moreover the kinetic terms at the  boundaries, when they are large and positive, 
just make some degrees of freedom more weakly coupled at the boundary
(When they are large and negative they lead to light tachyonic ghosts;
 see discussion below).
An explicit proof that boundary masses plus kinetic terms are innocuous could be obtained by using
the technique of ref.~\cite{lpr}. It would correspond to
choosing an analogue of 't Hooft-Feynman gauge where  both the $\sigma$-model Goldstones and the 
5th components of the vectors at the boundaries (which are normally set to zero when working on an orbifold)
are kept non-zero. In such a gauge, unlike in the unitary gauge,
 the  propagators are well behaved in the UV  and the loop power counting  straightforward.
\footnote{In ref.~\cite{cgmpt} 
it was shown that an infinite mass at the boundary does not lower the unitarity cut-off
of elastic  KK scattering. Inelastic processes were however not studied.}
Therefore, according to NDA, we define the strong coupling scales of the bulk gauge theories by
\beq
\Lambda_i\equiv 24 \pi^3 M_i\, ,\ \ \ \ i=L,R,B\, .
\eeq
Finally we assume the SM fermions to be (in first approximation) 
localized at $y=0$, away from the source of 
electroweak breaking. In this way, extra
unwanted non-oblique corrections are kept at a minimum. We will later comment on how to give fermions a mass while keeping the suppression of  extra four-fermion interactions.

\section{The low energy effective theory}

To study the low energy phenomenology one way to proceed is to find the KK masses and wave functions.
However, since the SM fermions couple to $W_L^I(x,y=0)\equiv {\bar W}^I(x)$ 
and $W_R^3(x,y=0)=
B(x,y=0)\equiv {\bar B}(x)$, 
it is convenient to treat the exchange of vectors in a two step procedure. First we integrate out the bulk
to obtain an effective lagrangian for ${\bar W}^I$ and $\bar B$. 
Then we consider the exchange of 
the interpolating fields ${\bar W}^I$ and $\bar B$ between light fermions.
Indeed we will not need to perform 
this second step: 
to compare with the data we  just need to extract the $\epsilon$'s \cite{alba},
or $S,T,U$ observables \cite{pt2},
from the effective lagrangian for the interpolating fields. This way of proceeding is clearly inspired by 
holography, though we do not want to emphasize this aspect for the time being.

To integrate out the bulk, we first must solve 
the 5D equations of motions, imposing at $y=\pi R$ the boundary conditions
that follow from the variation of the action, while at $y=0$ the fields are fixed at
$W_L^I(x,y=0)= {\bar W}^I(x)$ and $W_R^3(x,y=0)=B(x,y=0)={\bar B}(x)$. 
By substituting the result back into the action,
we obtain the  4D effective lagrangian
\beq
{\cal L}_{eff}=\int_0^{2\pi R}dy \left ({\cal L}_5+{\cal L}_\pi\right )\, .
\eeq
We will work in the unitary gauge gauge where the 5th components of the gauge fields are set to zero.
At the quadratic level in the 4D fields ${\bar W}_L^I(x), {\bar B}(x)$, integration by parts and use of the
equations of motion allows to write ${\cal L}_{eff}$ as a boundary integral
\beq
{\cal L}_{eff}= 
\left[M_L{{\bar W}^I}_{\mu}\partial_y W^{I\, \mu}_L
+\bB_\mu\left (M_R\partial_y W_R^{3\, \mu}+M_B\partial_yB^\mu\right)
\right ]_{y=0}\, .
\eeq
In order to solve the equations of motion it is useful 
to work in the momentum representation along the four
non-compact dimensions: $x_\mu \to p_\mu$. Moreover it is useful to separate the fields in the longitudinal and
transverse components $V_\mu(y)=V^t_\mu+V^l_\mu$ satisfying separate 5D equations
\bea
(\partial_y^2 -p^2)V_\mu^t&=&0\, ,\nonumber\\
\partial_y^2 V_\mu^l &=&0\, . 
\eea
Nevertheless,
 since we are interested in the coupling to light fermions, 
we will just focus on the
transverse part and eliminate the superscript alltogether. The general bulk solution has
the  form $V_\mu= a_\mu \cosh (py)+b_\mu \sinh (py)$ 
where $a_\mu$ and  $b_\mu$ are fixed by the
boundary conditions. After a straightforward computation we find
\beq
{\cal L}_{eff}= \bW^I_\mu \Sigma_{L}(p^2) \bW^{I\mu} +\bW^3_\mu \Sigma_{3B}(p^2)\bB^\mu+
\bB_\mu \Sigma_{BB}(p^2)\bB^\mu\, ,
\label{oblique}
\eeq
with
\bea
\Sigma_L&=&-M_L\frac{2M_L\, p\tanh (p\pi R)+2M_R\, p\coth (p \pi R) 
+Z_W\, p^2}{2(M_L+M_R)+Z_W\, p\tanh(p \pi R)}
\, ,\nonumber\\
\Sigma_{3B}&=&-\frac{4M_LM_R\, p (\tanh (p \pi R) -\coth (p \pi R)) }
{2(M_L+M_R)+Z_W\, p\tanh(p \pi R)}\, ,
\nonumber\\
\Sigma_{BB}&=&-M_R\frac{2M_R\, p\tanh (p \pi R)+2M_L\, p\coth (p \pi R) 
+Z_W\, p^2}{2(M_L+M_R)
+Z_W\, p\tanh(p\pi R)}\nonumber\\
&{}&-M_B\frac{2M_B\, p\tanh (p \pi R)+Z_B\, p^2}{2M_B+Z_B\, p\tanh(p \pi R)}\, .
\label{sigmas}
\eea
The total lagrangian
$
{\cal L}_0+{\cal L}_{eff}
$
gives us the complete effective theory as a function of the boundary fields at $y=0$.  
The KK spectrum of the model and the couplings of
the KK modes to the boundaries can be obtained
by finding the poles and residues of the full inverse kinetic matrix. 
It is  instructive (and also phenomenologically preferable, as we will discuss)
to consider the limiting case where the boundary  kinetic terms 
at $y=0$ dominate the contribution
from the bulk: $1/g^2,1/g'^2\gg M_i\pi R, Z_{W,B}$. 
In this limit the physical masses sit very close to the poles of the
$\Sigma/p^2$'s. For instance, for the charged vectors we have two towers of modes
that we call odd and even \footnote{Notice that when $M_L=M_R$ the bulk theory is invariant under parity
$W^L\leftrightarrow W^R$, 
and the odd and even modes are respectively vector and axial under parity.} .
In the limit $Z_W=0$ the odd modes are
\beq
m_{n+1/2}= \frac{n+\frac{1}{2}}{R}\left[1+\frac {2g^2 M_L^2 R}{(n+\frac{1}{2})^2(M_L+M_R)}+\dots\right]\ \ \ \ n=0,1,2,...
\label{oddmasses}
\eeq
and the even ones are
 \bea
m_{n}&= &\frac{n}{R}\left [1+\frac {2g^2 M_LM_R R}{\pi n^2(M_L+M_R)}+\dots\right ] \ \ \  \ n>0
\, ,\nonumber\\
m_0^2&=&\frac {2g^2 M_LM_R }{\pi (M_L+M_R)R}\left [1+O(g^2M_{L,R}R)\right ]\equiv m_W^2\, .
\label{wmass}
 \eea
In the limit we are considering, 
the lightest mode 
is much lighter than the others $m_0\ll 1/R$: it sits
close to the Goldstone pole of $\Sigma_L/p^2$. 
This mode should be interpreted as the usual $W$-boson of the
standard model. 
Similarly in the neutral sector we find a lightest massive boson, the $Z$, with mass
\beq
m_Z^2=\frac {2(g^2+g'^2) M_LM_R }{\pi (M_L+M_R)R}\left [1+O(g^2M_{L,R}R)\right ]\, .
\label{zmass}
\eeq
One can compute the couplings of $Z$, $W$ and photon to elementary fermions from the first
$\partial_{p^2}$ derivative of the full 2-point function at the zeroes. 
For small $g^2M_{L,R}R$,
one finds that  the leading contribution 
arises from the boundary lagrangian ${\cal L}_0$. 
We then conclude that, up to corrections of  ${\cal O}(g^2M_{L,R}R)$, 
the SM  relations between masses and couplings are satisfied.
This can also be seen from the wave-function
of the lowest  KK modes that,
for large kinetic terms on the $y=0$ boundary, 
is peaked at $y=0$,
and then the SM gauge fields
correspond approximately to the boundary fields $\bW^I$ and $\bB$.

All the deviation from the SM at the tree level 
are due to oblique corrections and can be studied
by expanding eq.~(\ref{sigmas}) at second order in $p^2$ around $p^2=0$:
$\Sigma(p^2)\simeq\Sigma(0)+p^2\Sigma^\prime(0)$. As custodial isospin is manifestly preserved
by eq.~(\ref{oblique}),
 we find that the only non-zero observable is $\epsilon_3$ ($S$
parameter):
\beq
\epsilon_3=-g^2\Sigma^\prime_{3B}(0)=
g^2\frac{4\pi R}{3}\frac{M_LM_R}{M_L+M_R}\left [1+\frac{3 Z_W}{4\pi (M_L+M_R)R}\right ]\, .
\label{eps3}
\eeq
It is convenient to define $1/\Lambda=1/\Lambda_R+1/\Lambda_L$, so that $\Lambda$ is 
essentially the cut-off of
the theory. It is also convenient to define $Z_W\equiv \delta/16\pi^2$, as $\delta=O(1)$ corresponds
to the natural NDA minimal size of boundary terms. This way $\epsilon_3$ is rewritten as
\beq
\epsilon_3=\frac{g^2}{18\pi^2}(\Lambda R)\left [ 1+ \frac{9\delta}{8 (\Lambda_R+\Lambda_L)R}\right ]\, .
\label{result}
\eeq
Now, the loop expansion parameter of our theory is $\sim 1/(\Lambda R)$. In order for all our approach to be 
any better that just a random strongly coupled electroweak breaking sector we need $\Lambda R\gg 1$.
This can be reconciled with the experimental bound $\epsilon_3 \lsim 3 \cdot 10^{-3}$ \footnote{$\epsilon_3 \lsim 3 \cdot 10^{-3}$ is the limit on the extra contribution to $\epsilon_3$ relative to the SM one with $m_H=115$ GeV. 
The limit is at 99$\%$ C.L. for a fit with $\epsilon_1$ and $\epsilon_3$ free, 
but $\epsilon_2$ and $\epsilon_b$ fixed at their SM values. 
Letting $\epsilon_2$ and $\epsilon_b$  be also
free would weaken the limit, but only in a totally marginal way.}
only if
$\delta$ is large and negative, $\delta \sim -\Lambda R$, and partially cancels the leading term
in  eq.~(\ref{result}).
This requirement, however, leads to the presence of a tachyonic ghost 
with $m^2\sim -1/R^2$ in the vector spectrum, again a situation that would make our effective theory useless.

\section{Fermion masses}

The discussion so far assumed the SM fermions
to be exactly localized at $y=0$, in which case they would be exactly massless, having no access to the electroweak
breaking source at $y=\pi R$. As we have argued, in this limiting case there are only oblique corrections to
fermion interactions: all the information that there exists 
an extra dimension (and some strong dynamics)
is encoded in the vector self-energies in eq.~(\ref{oblique}). 
For example, there are no additional 4-fermion
contact interactions. 

A more realistic realization of fermions is the following. 
Consider bulk fermions with
the usual quantum numbers under $SU(2)_L\times SU(2)_R\times U(1)_{B-L}$. 
For instance,    the right handed fermions
will sit in $(1,2, B-L)$ doublets. Each such 5D fermion upon orbifold
 projection
will give rise to one chiral multiplet with the proper quantum numbers under $SU(2)_L\times U(1)_Y$.  Fermion mass operators mixing left and right multiplet can then be written at
$y=\pi R$, very much like the $W_R-W_L$ vectors. However in the case of fermions, since their mass dimension 
in 5D is 2, the mass coefficient at the boundary is dimension-less. Then, in the limit in which the scale of
electroweak breaking at $y=\pi R$ is sent to infinity, the fermion mass at the boundary should stay finite.
Indeed if we indicate by $F$ the scale of electroweak breaking at the boundary and put back the Goldstone field matrix $U$
which non linearly realizes the symmetry, the fermion mass operator has the form $\bar \psi U\psi$ while
the vector mass arises from the Goldstone kinetic term $F^2(D_\mu U)^\dagger (D^\mu U)$. 

Finally,
to make the masses small, to break isospin symmetry, and to cause
effective approximate localization on the $y=0$ boundary it is  enough to add large kinetic terms for the 
fermions at $y=0$. Since isospin is broken at $y=0$ the kinetic terms will distinguish fermions
with different isospin, for instance $s_R$ from $c_R$, and a realistic theory
can be easily obtained.
Fermion masses $m_f$ will go roughly like $1/\sqrt{Z_L Z_R}$, where $Z_L$ and $Z_R$
are the boundary kinetic coefficients of the left and right handed components respectively.
Notice that the $Z$'s for fermions have dimension of length. Non oblique 
effects due to the
fermion tail into the bulk will then scale like $\sim R/Z\sim m_f/m_{KK}$, which is negligible for the 2 light 
generations, but probably not for the bottom quark. One could also spread the fermions more into the bulk
by decreasing the $Z's$ while decreasing at the same time the fermion mass coefficients at the 
electroweak breaking boundary.
This way one would get more sizeable non-oblique corrections to EWPT, in the form of corrections
to the $W$ and $Z$ vertices and four-fermion interactions from the direct coupling to vector KK modes. In principle
these extra parameters could be used to improve the fit, tuning the effective $\epsilon_3$ small.
We do not pursue this here, since we do not think of  this possibility as being very compelling.

\section{$\eps_{3}$ 
for  a general metric}

One can wonder whether different 5D geometries can change the result above
by giving, for example,
a negative contribution to $\eps_{3}$.
Here we will  show that this is not the case and 
$\eps_3$ stays positive whatever the metric. 
As we want to preserve 4D 
Poincar{\'e} symmetry the  
curvature will just reduce to a warping. It is convenient to choose the 5th coordinate $y$ in such
a way that the metric is
\beq
ds^2=e^{2\sigma(y)}dx^\mu dx_\mu +e^{4\sigma(y)} dy^2\, ,
\eeq
and take, to simplify the notation, $0\leq y \leq 1$ ($\pi R=1$).
With this choice the bulk equation of motion of a transverse vector field becomes
\beq
(\partial_y^2-p^2 e^{2\sigma})V_\mu=0\, .
\label{warpedeom}
\eeq
To simplify the discussion we  limit ourselves to the case $M_R=M_L=M$, in which parity is conserved in the bulk and we set $Z_W=0$.
Moreover, as $U(1)_{B-L}$ does not play any role in $\eps_3$, 
we neglect it alltogether.
 The extension to the general case 
is straightforward. 

To calculate $\eps_3$ it is  convenient to work in the basis of vector 
$V=(W_L+W_R)/\sqrt 2$ and axial 
$A=(W_L-W_R)/\sqrt 2$ fields, for which 
\beq
\eps_3 = -\frac{g^2}{4}\big[\Sigma^\prime_{V}(0)-\Sigma^\prime_{A}(0)\big]\, ,
\label{vma}
\eeq
where $\Sigma_{V}=4MV^{(-1)}\partial_y V|_{y=0}$ and similarly for $\Sigma_{A}$.
Since we are only interested in the $\Sigma$'s 
at $O(p^2)$, we can write the solutions of the bulk equations of motion
as 
\bea
V_\mu&=&\bar V_\mu(p^2)\left (v^{(0)}(y)+p^2 v^{(1)}(y)+O(p^4)\right )\, ,\nonumber\\
A_\mu&=&\bar A_\mu(p^2)\left (a^{(0)}(y)+p^2 a^{(1)}(y)+O(p^4)\right )\, ,
\eea
where, as before, $\bar A_\mu$ and $\bar V_\mu$ are the fields at $y=0$. 
The functions $v^{(0)}$ and $a^{(0)}$ solve eq.~(\ref{warpedeom})
at $p^2=0$, with boundary conditions $v^{(0)}(0)=a^{(0)}(0)=1$ and 
$\partial_y v^{(0)}|_{y=1}=a^{(0)}(1)=0$. We obtain
\beq
v^{(0)}=1\, ,\quad\quad a^{(0)}=1-y\, .
\eeq
The function $v^{(1)}$
solves
$
\partial_y^2 v^{(1)}= e^{2\sigma(y)} v^{(0)}
$
with boundary conditions $v^{(1)}(0)=\partial_y v^{(1)}|_{y=1}=0$.
For  $a^{(1)}$ the boundary conditions are  $a^{(1)}(0)= a^{(1)}(1)=0$.
Then we find
\beq
\epsilon_3=
g^2 M\left \{ \int_0^1 e^{2\sigma(y)} dy\,-\,\int_0^1dy\int_0^y(1-y')e^{2\sigma(y')} dy'\right \}\, ,
\label{positiveS}
\eeq
which is manifestly positive. The first term can in fact be written as an integral in two variables
by multiplying by $1=\int_0^1 dy'$. Then, since $e^{2\sigma}>(1-y)e^{2\sigma}>0$ 
and  the domain of integration 
of the second term is a subset of the domain of the first, $\eps_3>0$ follows.

While $\eps_3$ is always positive, it could become very small if $e^{2\sigma}$ decreases rapidly away from $y=0$.
However this is the situation in which the bulk curvature becomes large. Indeed the Ricci scalar is given by
${\cal R}=[8\sigma''+4(\sigma')^2]e^{-4\sigma}$ and tends to grow when the warp factor decreases. In fact the curvature length
scales roughly like $e^{2\sigma}$, so that the general expression in
eq.~(\ref{positiveS}) roughly corresponds
to the flat case result, eq.~(\ref{result}), with the radius $R$ replaced by the curvature length at $y=1$.

As an explicit example consider the family of metrics (in the conformal frame)
\beq
ds^2=(1+y/L)^{2d}\left \{ dx_\mu dx^\mu+dy^2\right \}\, ,
\eeq
with $0\leq y\leq \pi R$ as originally. With this parametrization the massive KK modes are still quantized in units
of $1/R$. On the other hand the curvature goes like 
\beq
{\cal }{\cal R}\sim \frac{1}{L^2 (1+y/L)^{2+2d}}\, ,
\eeq
so that for $d<-1$ it grows at the electroweak breaking boundary ($d=-1$ corresponds to AdS). Indicating $\Delta=(1+\pi R/L)$
we have
\beq
\eps_3= g^2 M\frac{L}{d+1}\left \{-1+\Delta^{1+d}+
\frac{4+(d-1)^2(\Delta^2-1)+(d-3)\Delta^{1-d}-(1+d)\Delta^{d-1}}{(3-d)(2-\Delta^{1-d}-\Delta^{d-1})}\right \}\, ,
\eeq
which for $d<-1$, $L\ll R$ (big warping) 
gives roughly  $\eps_3\simeq g^2\Lambda R_C/(18\pi^2)$
where the curvature length at $y=1$, $R_C\sim L\Delta^{d+1}$, 
has replaced the radius $R$.
Of course when $\Lambda R_C<1$ 
we loose control of the derivative expansion 
for our gauge theory. 
We then conclude that the  model in the  regime of calculability, $\Lambda R
,\Lambda R_C\gg 1$ , gives always large contributions to $\epsilon_3$.

\section{Similarities with  strongly coupled 4D theories}

In this Section we comment on the relation between the model presented 
here and technicolor-like theories 
 in 4D. We will consider the case $M=M_L=M_R$ and $Z_W=0$.

As done for the case of AdS/CFT
one can establish a qualitative correspondence between the bulk theory (in fact the bulk plus the $y=\pi R$ boundary)
and a purely four dimensional field theory with a large number of particles, indeed
a large $N$ theory. The loop expansion parameter $1/(\Lambda R)$ of the 5D theory corresponds
to the topological expansion parameter $1/N$, so that our tree level calculation corresponds
to the planar limit. Consistent with this interpretation, the couplings among the individual KK bosons
go like $1/\sqrt N$, as expected in a large $N$ theory. 
The result  of eq.~(\ref{result}), written as  $\eps_3\sim g^2 N/18\pi^2$, 
also respects the correspondence: it looks precisely like what one would expect in a large $N$ technicolor.
Similarly the 1-loop gauge 
contribution to $\eps_{1}$ and $\eps_{2}$ is proportional to $g'^2/16\pi^2$, 
with no $N$ enhancement. This result is easy to understand in the ''dual'' 4D theory. Custodial isospin is 
only broken by a weak gauging of hypercharge by an external gauge boson ($\equiv$ living at the $y=0$ boundary).
Isospin breaking loop effects involve
the exchange of this single gauge boson and are thus not enhanced by $N$. 
Similar considerations can be made for the top contribution.
Of course, though useful,
this correspondence is only qualitative, in that we do not know the microscopic theory on the 4D side.
We stress that this qualitative correspondence is valid whatever the metric of the 5D theory. The case of
AdS geometry only adds conformal symmetry into the game allowing for an (easy) extrapolation (for a subset of
observables) to arbitrarily high energy. On the other hand, when looking for solutions to the little hierarchy 
problem \cite{rb}, one can be content with a theory with a fairly low cut-off (maybe 10 TeV) in which case
conformal symmetry is not essential.

The correspondence with a large $N$ technicolor also goes through for the sign of $\eps_3$. Very much like 
 $S$ is positive in rescaled versions of QCD \cite{tech,pt1}, 
we have proven a positive $S$ theorem for a class
of ``holographic'' technicolor theories. 
Is there a simple reason for this relation? Perhaps some
insight can be obtained by realizing that the $\Sigma_{V,A}$, both in our 5D models and in a
generic large $N$ theory, can be rewritten as a sum over narrow resonances
 \beq
\Sigma_V=-p^2\sum_n\frac{F^2_{V_n}}{p^2+m^2_{V_n}}\, ,\ \ \ \ \
\Sigma_A=-p^2\sum_n\frac{F^2_{A_n}}{p^2+m^2_{A_n}}-f^2_\pi\, .\ \ \ \ \
\label{valargen}
\eeq
Then, from eq.~(\ref{vma}) we have 
\beq
\eps_3=\frac{g^2}{4}\sum_n\left[\frac{F^2_{V_n}}{m^2_{V_n}}-
\frac{F^2_{A_n}}{m^2_{A_n}}\right]\, .
\label{eps3largen}
\eeq
For a flat extra dimension we have 
$F^2_{V_n}=F^2_{A_n}=8M/(\pi R)$, $f^2_\pi=4M/(\pi R)$,
and the masses
  $m_{V_n}=(n+1/2)/R$ and  $m_{A_n}=(n+1)/R$ with $n=0,1,2,...$.
Then $\eps_3$ is dominated by the first resonance,
a vector, and turns out positive. 
Qualitatively, what happens is the following: 
vector and axial resonances alternate
in the spectrum and, since the lightest state 
is a massless Goldstone boson in the axial channel, the lightest
massive state tends to be a vector, so that $\eps_3$ tends to be positive.

In ref. \cite{pt1} positive $\eps_3$ was deduced 
by  saturating eqs.~(\ref{valargen}) and ~(\ref{eps3largen}) with the first two low
laying $J=1$ resonances, called $\rho$ and  $a_1$ mesons,  
after imposing the two Weinberg sum rules \cite{sr}:
\beq
\epsilon_3^{TC}= \frac{g^2}{4}\left(1+\frac{m^2_\rho}{m^2_{a_1}}\right)\frac{f^2_\pi}{m^2_\rho}\, .
\label{largen}
\eeq
In our 5D model 
 $\Sigma_V-\Sigma_A$ vanishes exponentially 
in the large euclidean momentum region, 
and then  
an infinite set of generalized Weinberg sum rules are satisfied
\footnote{These arise by imposing that the coefficients of 
the Taylor expansion of $\Sigma_V-\Sigma_A$ (from eq.~(\ref{valargen}))
 at $p^2\rightarrow\infty$ are zero.}.
However all levels are involved in the sum rules, so that,
strictly speaking, one cannot play rigorously the same game. 
There are, nevertheless, some surprising 
numerical coincidences.
For example, in flat 5D, we have 
\beq
m_\rho=\frac{1}{2R}\, ,\ \ \ \ 
m_{a_1}=\frac{1}{R}\, ,
\eeq
and eq.~(\ref{eps3}) can be rewritten as 
\beq
\epsilon_3= \frac{g^2\pi^2}{30}\left(1+\frac{m^2_\rho}{m^2_{a_1}}\right)
\frac{f^2_\pi}{m^2_\rho}\, .
\label{result2}
\eeq
This result deviates by less than a  30$\%$ from the expression of 
eq.~(\ref{largen}).

Based on these considerations, one is therefore driven to establish a connection, inspired by holography, between the 5D
 model presented here and 
technicolor-like  theories in the large $N$ limit. If this is the case, the impossibility to fit the EWPT, while keeping 
the perturbative expansion, goes in the same direction as the claimed difficulty encountered in 4D 
technicolor-like  theories to account for the EWPT.

\section*{Acknowledgements}
\label{sec:acknowledge}

We would like to thank Kaustubh Agashe, Nima Arkani-Hamed, Roberto Contino, Csaba Cs{\'a}ki, Antonio Delgado, Paolo Gambino,
Tony Gherghetta, Gian Giudice, Christophe Grojean, Markus Luty, Santi Peris,
Luigi Pilo, Raman Sundrum and John Terning
for very useful discussions.  AP and RR thank the Physics Department at Johns Hopkins University for hospitality.
RR thanks the Aspen Center for Physics and the participants to the workshop ``Theory and Phenomenology at the Weak Scale''.
This work has been partially supported by MIUR and by the EU under TMR contract
HPRN-CT-2000-00148.
The work of AP was  supported  in part by 
the MCyT and FEDER Research Project
FPA2002-00748 and DURSI Research Project 2001-SGR-00188.

\end{document}